# Not All Qubits Are Created Equal
## A Case for Variability-Aware Policies for NISQ-Era Quantum Computers

Swamit S. Tannu
Georgia Institute of Technology
swamit@gatech.edu

Moinuddin K. Qureshi
Georgia Institute of Technology
moin@ece.gatech.edu

*Abstract*—Recently, IBM, Google, and Intel showcased quantum computers ranging from 49 to 72 qubits. While these systems represent a significant milestone in the advancement of quantum computing, existing and near-term quantum computers are not yet large enough to fully support quantum error-correction. Such systems with few tens to few hundreds of qubits are termed as *Noisy Intermediate Scale Quantum computers (NISQ)* and these systems can provide benefits for a class of quantum algorithms. In this paper, we study the problems of *Qubit-Allocation* (mapping of program qubits to machine qubits) and *Qubit-Movement* (routing qubits from one location to another to perform entanglement).

We observe that there exists variation in the error rates of different qubits and links, which can have an impact on the decisions for qubit movement and qubit allocation. We analyze characterization data for the IBM-Q20 quantum computer gathered over 52 days to understand and quantify the variation in the error-rates, and find that there is indeed significant variability in the error rates of the qubits and the links connecting them. We define reliability metrics for NISQ computers and show that the device variability has significant impact on the overall system reliability. To exploit the variability in error rate, we propose *Variation-Aware Qubit Movement (VQM)* and *Variation-Aware Qubit Allocation (VQA)*, policies that optimize the movement and allocation of qubits to avoid the weaker qubits and links, and guide more operations towards the stronger qubits and links. We show that our Variation-Aware policies improves the reliability of the NISQ system upto 2.5x .

## I. INTRODUCTION

Quantum computers can accelerate conventionally hard problems such as prime-factorization, understanding photosynthesis, and simulation of materials and molecules [1], [2]. Quantum algorithms use quantum bits (qubits) to exploit the properties of superposition and entanglement, and rely on quantum operations to change the state of the qubits. In the last two decades, the field of quantum computing has moved from theoretical ideas to realizable systems (albeit at a small scale). The last two years represent significant milestones in the field of quantum computing, as Google, IBM, and Intel have announced quantum computers with 72, 50 and 49 qubits respectively [3], [4], [5]. Figure 1 shows some of the recent quantum machines. The availability of quantum computers provide an opportunity for system designers and architects to understand the problems and challenges in building and operating a realistic quantum computer and use these insights to guide the design of future larger-scale quantum computers.

Qubit devices can lose state due to decoherence, and the operations on qubits can also experience errors. Qubits can be protected against errors using specialized codes, called *Quantum error correction codes (QEC)*. Unfortunately, QEC requires significant overheads, typically incurring 10-50 physical qubits to encode one fault-tolerant qubit. Existing and near-term quantum computers with tens to hundreds of qubits will not have the capacity to utilize QEC due to the limited number of qubits. Such quantum computers with 10 to 1000 qubits, operating in noisy environments are termed as *Noisy Intermediate Scale Quantum computers (NISQ)* [6]. Even though NISQ machines may not have enough resources for error correction, they can still provide significant benefits for a class of quantum applications. In this paper, we study policies for Qubit-Movement (routing a data qubit from one location of the chip to another) and Qubit-Allocation (mapping of program qubits to the physical qubits) for NISQ machines.

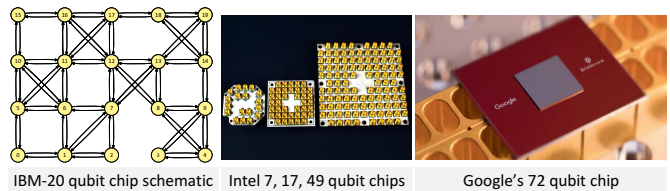

IBM-20 qubit chip schematic | Intel 7, 17, 49 qubit chips | Google's 72 qubit chip

**Fig. 1:** Recent demonstrations of Quantum Computers [3], [5], [4]

Power of quantum computers come from the ability to generate a collective entangled state. An entangled state is generated by coupling a pair of qubits using two qubit operation. A machine can entangle only the qubits that have a link between them. Existing solid state quantum computers from IBM, Google, and Intel, are designed using networks that offer limited connectivity, only to a few of the neighboring qubits, and this connectivity dictates the qubits that can be entangled. For example, Figure 2(a) shows a hypothetical quantum computer with five qubits where circular nodes represent the qubits and edges represent the coupling links between qubits. A pair of qubits can only be entangled if there exists a coupling link between them. Fortunately, quantum computers provide a *SWAP* instruction that can exchange the state of two neighboring qubits. For example, we want to entangle data qubit $Q_1$ and data qubit $Q_3$ which are initially residing at physical qubit-A, and physical qubit-C respectively. We can perform this operation in two steps: first swap the data between qubit-A and qubit-B such that $Q_1$ and $Q_2$ interchanges positions. Next, entangle qubit data $Q_1$ and $Q_3$. In quantum programs, large number of SWAP instructions are inserted to move data so that entanglement between arbitrary qubits can be performed. The insertion of SWAP instructions is done statically by a compiler, therefore the information about link usage is available and deterministic [7], and routing can be done without deadlocks.

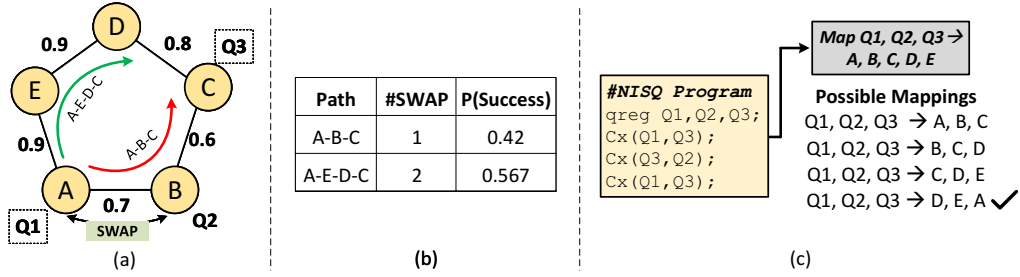

**Fig. 2:** (a) A hypothetical quantum computer with five-qubits – the number on the edge denotes the success probability when that edge is used (b) Variation-Aware Qubit Mapping (VQM) can use more SWAP instructions and yet have higher probability of success (c) Variation-Aware Qubit Allocation (VQA) tries to select the mapping that improves overall system reliability.

The *Qubit-Movement* policy deals with the problem of selecting a route to move the data of one qubit to another. For example, in Figure 2(a), we may choose the route A-B-C for going from A to C, as doing so would minimize the number of SWAP operations. The *Qubit-Allocation* policy deals with the problem of mapping of program qubits to the physical qubits. For example, in Figure 2(a), if we want to map three program qubits to five physical qubits, we would choose any of 3 connected qubits (for example, $Q_1$ maps to A, $Q_2$ maps to B, and $Q_3$ maps to C), as placing qubits nearby results in efficient movement. Recent studies [8], [9], [10] have proposed qubit allocation policies based on minimizing the number of SWAP instructions. These studies assume uniformity in cost of performing SWAP operations. However, in reality, we expect variation in the behavior of different qubits and links, and optimizing for a uniform behavior may not result in the best policy when device variation is taken into account.

To understand and quantify the variation in the error-rates of different qubits and links, we analyze the publicly-available characterization data for the IBM-Q20 (20 qubit) machine. Such characterization is performed for the IBM-Q20 several times a day, and we analyze the data for a period of 52 days. We present the statistics of coherence time for all the 20 qubits, the error rate in performing single-qubit operations, and the error-rate in performing two-qubit operations across different qubits. For all these metrics we observe significant variation in the behavior of different qubits and links – in essence, qubits and links are not created equal. For example, our detailed analysis for the links connecting different qubits show that the error rates can vary by as much as 7x across different links in the system. Such variation can have a significant impact on the overall system reliability (Section III).

To analyze the impact of variation on the reliability of NISQ machines, we develop two system-level reliability metrics: *Mean Instructions Before Failure (MIBF)* and *Probability of Successful Trial (PST)*. For programs that are long-running and have negligible probability of completion without a failure, MIBF denotes the amount of operations performed before the first error is encountered. For programs that tend to finish successfully some of the times, the PST metric indicates the probability that the program finished successfully without any error. We build an evaluation infrastructure to compute the MIBF and PST for the IBM-Q20 machine, and performed analysis using small applications and kernels. Our analysis shows that the device variation has a significant impact on the system level reliability. To improve system reliability, we should steer more instructions and movement to strong qubits and links, and fewer instructions and movement on weaker qubits and links. We propose such *Variation-Aware* policies to exploit the variation in the behavior of qubits and links, assuming that device level characterization data is available.

We propose *Variation-Aware Qubit Movement (VQM)* policy that routes the qubit from source to destination based on minimizing the probability of failure. For example, in Figure 2, the success probability of each link is denoted as a weight of the edge. Let us assume, we want to entangle data qubit $Q_1$ and data qubit $Q_3$. A conventional variation-unaware policy will use a path that minimizes the number of SWAP instructions, taking the path A-B-C, resulting in an overall probability of success of 42% for these operations. With VQM, we would take the route A-E-D-C, even though this route has more SWAP instructions, since it has an overall probability of success of 56.7%, as shown in Figure 2(b). We make VQM tunable with a parameter that limits the number of extra SWAP instructions allowed on a route. Our evaluations show that VQM improves MIBF upto 1.5x.

We also propose *Variation-Aware Qubit Allocation (VQA)* policy that performs the mapping of program-qubit to physical-qubit with an aim of improving overall system reliability. For example, in Figure 2(c), we want to do an allocation of three program qubits to 5 physical qubits. A conventional mapping policy can choose any of the listed mapping possibilities as they all would have similar cost in terms of SWAP operations. However, with VQA, we would use the mapping D, E, A, as this mapping uses the strongest links, and would improve the overall system reliability. We extend prior proposals for Qubit-Allocation with VQA, and show that VQA improves system reliability significantly.

We also perform a case study, where we analyze programs that require less than half the available qubits and we have an option of either executing two copies of the same NISQ program concurrently (to increase the rate at which trials are performed) or executing only one copy but mapping the work on strongest qubits and links (to improve the PST). We demonstrate that, in certain scenarios, operating one strong copy has better overall performance (successful trials per unit time) than running two copies. Thus, variation-awareness can influence intelligent partitioning policies for NISQ computers.

## II. BACKGROUND AND MOTIVATION

In this section, we provide a brief background on quantum computing, discuss the issues of errors and error correction, and present a usage model for NISQ computing and the problems associated with the NISQ model.

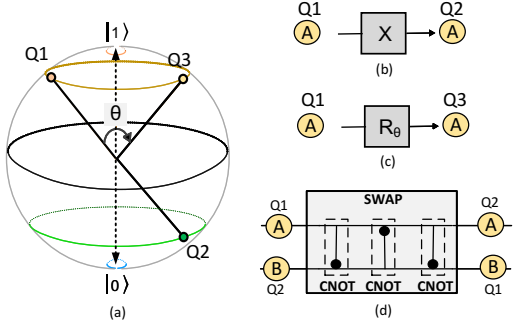

**Fig. 3:** (a) Bloch Sphere representation of qubit. (b)–(c) Quantum operations manipulate the state by moving the point on sphere. (d) SWAP instruction interchanges the qubit-data between two devices and can be accomplished using 3 CNOT operations.

### A. Background on Quantum Computing

Conventional computers use binary data representation. Whereas, a quantum computer represents data using quantum bits (qubits). Consider a sphere, where the binary data can either be at the north-pole or the south-pole of the sphere, and conventional digital computers operate by switching the data between the north and south poles. In quantum computing, the state of a qubit can be viewed as any arbitrary point on the sphere that is a superposition of two basis states as shown in Figure 3(a). Quantum operations manipulate the state of the qubit by moving it from one point to another point on the sphere, as shown in Figure 3(b) and Figure 3(c). The ability to store and manipulate the state of qubits enables efficient quantum algorithms.

The second property that facilitates quantum parallelism is *entanglement*. Entanglement is the ability to produce a collective state of multiple qubits that are correlated, and manipulating one qubit can have an impact on the state of the other qubit(s). The entangled states are produced using a series of two qubit operations such as the *Controlled-NOT (CNOT)* instructions [11].

On IBM quantum machines, two-qubit operations are performed using a coupling-link that connects two qubits. For practical reasons, superconducting quantum computers do not allow all-to-all connectivity between the qubits, and use a restricted network (such as Mesh) that allows connectivity between only the neighboring qubits. The network structures impose constraints on which qubits can be entangled. Fortunately, there are SWAP operations that can move the qubit from one location to another, and enables entanglement of any two arbitrary qubits. Even if the quantum machine does not provide a native SWAP instruction, such an operation can be accomplished using 3 CNOT gates, as shown in Figure 3(d).

### B. Errors in Quantum Computers

Qubits are fickle as even a small perturbation in the environment can change the state of a qubit. Error rate for a qubit can be defined as probability of undesired change in the qubit state. Errors in quantum computers can be classified into two categories: retention-errors or operational-errors.

**Retention Errors (or Coherence Errors):** A qubit can retain data for only a limited time, and this duration is called as *Coherence Time*. There are two types of retention errors that can occur, and there are two metrics to specify the coherence time of a quantum device. A qubit in an high-energy state (state $|1\rangle$) naturally decays to the low-energy state (state $|0\rangle$), and the time constant associated with this decay is called as the *T1 Coherence Time*. T1 indicates the time for natural relaxation of qubit (an architectural analogy would be the average retention time of DRAM cells). However, there is also a possibility that qubit might interact with environment and encounter a phase error even before relaxing into $|0\rangle$ state, and the time constant associated with this decay is called the *T2 Coherence Time*. T2 indicates the time for a qubit to get affected by the environment (an architectural analogy would be the average time for a DRAM cell to get flipped by a transient error).

The coherence times for superconducting quantum computers have improved from 1 nano-second to 100 micro-seconds in last decade [12]. Furthermore, existing superconducting qubits show improving trend in coherence times [4][12].

**Operational Errors (or Gate Errors):** Performing operations on qubits can also affect their state incorrectly due to errors, as quantum operations are not perfect. For example, an instruction that rotates the state by some desired angle can introduce extra erroneous rotation. Operational error-rate is defined as the probability of introducing an error while performing the operation [13]. For publicly available quantum-computers from IBM, the single-qubit instruction error-rates are of the order of $10^{-3}$, whereas for two-qubit instructions, such as CNOT, it is $10^{-2}$. Google Quantum machine [4] is reported to have about one-order of magnitude lower error rates than the IBM machines, however, detailed characterization data for this machine is not publicly available. A typical quantum program contains significant number of two-qubit operations, and given the error-rate of two-qubit operations are an order of magnitude higher than for the single-qubit operations, the overall error rate is usually dominated by the two-qubit operations. In this paper, we focus on operational errors, and specifically the ones caused by two-qubit operations.

### C. Quantum Error Correction and Overheads

Quantum computers can be made resilient to errors by using *Quantum Error Correction (QEC)* codes. Unfortunately, QEC requires a large number of physical qubits (10x-50x) to encode one fault-tolerant bit. This 10x-50x overhead in terms of physical qubits for performing error correction may be acceptable when the quantum machines have thousands of qubits, however, the current and near term quantum machines will not have enough capacity to implement error correction.

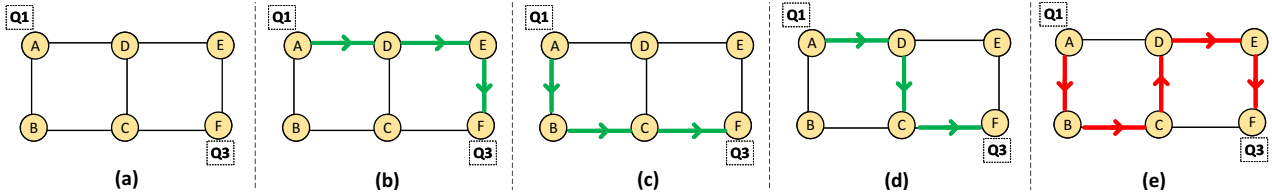

**Fig. 4:** (a) Layout of a 6-qubit quantum computer, (b)-(e) are possible routes from A to F. Note that options (b)(c)(d) have identical number of swaps and (e) incurs higher swaps. An intelligent policy would choose one from (b)(c)(d).

### D. Noisy Intermediate-Scale Quantum Computing (NISQ)

Executing large-scale quantum application, such as Shor's factoring algorithm, requires having a quantum computer with millions of qubits. Existing quantum technologies are not mature enough to have millions of qubits. In fact, for existing quantum computers (fifty-plus qubits) or near-term quantum computers (with few hundreds of qubits), it may be impractical to perform error correction even for an application requiring few dozen of logical qubits. However, there exists a class of applications highlighted by Preskill [6] that can still be viable with such *Noisy and Intermediate-Scale Quantum (NISQ) computing*. Even though NISQ machines may not have enough resources for error correction, they rely on application properties to perform useful work. To the best of our knowledge this is the first paper to investigate the architecture and operation for NISQ computers. Therefore, we first describe possible models of computing with the NISQ machines (alternative models for using NISQ machines are also possible).

If we had a program performing only a few qubit operations, then we could run the NISQ program through the NISQ computer and measure the qubit states, as shown in Figure 5(a). However, in such a model we would not know if the application encountered an error or not, given that the NISQ machines does not perform error correction. If there is a way to check if the output is correct or not algorithmically (for example, multiplying factors to see if the number can be obtained etc.) then we can rely on the output. However, a more general model for large number of operations is to run the program multiple times and log the output in each trial. As long as the correct results appear with non-negligible probability, we may be able to learn the correct results by analyzing the log.

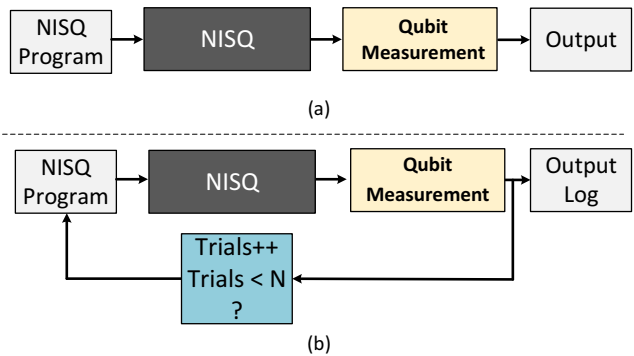

**Fig. 5:** Models for NISQ: (a) single-shot model (b) iterative model

### E. Problem: Restricted Connectivity Between Qubits

In this paper, we focus on the problems due to the architecture of NISQ computers. If a NISQ computer contains $N$ qubits, then ideally all the qubits will be connected to all other qubits. Such an unrestricted connectivity would allow any two arbitrary qubits to get entangled. Unfortunately, such an organization would require approximately $(N^2)$ links, which is impractical even for the 49-72 qubits machines that are available today. The links in a quantum machine are not just wires, but resonators that operate at dedicated frequency, and having a large number of such circuits operate reliably on the chip is a difficult task. Therefore, almost all qubit machines use either a Mesh network (or a variant that allows diagonal connections). Such networks restrict that the movement of qubits can occur only between neighboring qubits. For example, for the hypothetical 6-qubit machine shown in Figure 4(a) there is no direct connection between qubits A and F. The communication between these qubits must happen via intermediate qubits. Such restrictions give rise to the two sub-problems that we analyze: (a) Qubit-Movement policy, and (b) Qubit-Allocation policy.

**Qubit-Movement Policy:** This policy decides the route that should be used while moving the data from one location on the chip to another. Given that such movement is done using SWAP instructions between neighboring qubits, it is reasonable to select the route that minimizes the number of SWAP instructions. Figure 4(b)-(e) shows the four possible routes from A to F. The first three (b)-(d) requires only 3 SWAP operations, while (e) requires 4 SWAP operations. The policy may arbitrarily pick one of the routes from (b)-(d).

**Qubit-Allocation Policy:** This policy decides the initial mapping of program qubits to the data qubits. For example, it is preferred that qubits that communicate frequently be placed near each other. For example, if we wanted to place 4 qubits on the machine shown in Figure 4(a), we would not keep these qubits on the four corners, and instead we will try to use the middle two qubits (D and E), as doing so would minimize the SWAP instructions, required for communication. In fact, recent studies [8], [9], [10] have proposed such allocation policies based on minimizing the number of SWAP instructions.

Existing policies for Qubit-Movement and Qubit-Allocation assume uniform cost (specifically reliability impact) in performing SWAP operations. However, in reality, there can be significant variation in reliability of qubits and the links. Policies that take this variation into account can provide better overall system behavior (performance, reliability etc.) To enable such variation-aware policies, we first analyze the characterization data for the IBM-Q20 machine.

## III. ANALYZING VARIATION IN IBM-Q20

To understand and quantify the variation in the error-rates of different qubits and links, we analyze the publicly-available characterization data for the IBM-Q20 (20-qubit) machine. The data for link error rates and the coherence times gets published on IBM quantum experience web-page [14]. We monitored the IBM website for 52 days and gathered more than 50 different characterization reports. The characterization reports consist of error-rate for all single-qubit operations, two-qubit operations (link errors), and measurement operations. Furthermore, report contains the T1 and T2 coherence time for all 20 qubits. IBM machines are calibrated almost every day and error-reports are updated after each calibration cycle.

### A. Distribution of Coherence Times

Both T1 and T2 coherence time of a qubit depends on several design, manufacturing and experimental parameters. Due to process variation, biasing and temperature drifts the coherence time can vary significantly. Figure 6 shows the T1 and T2 distribution of IBM-Q20. The data is collected for all 20 qubits over 50 plus observations. T1 coherence data shows wider spread as compared to the T2 coherence time. The mean and standard deviation for T1-Coherence time are $80.32\mu S$ and $35.23\mu S$ respectively. And, the mean and standard deviation for T2-Coherence time are $42.13\mu S$ and $13.34\mu S$ respectively.

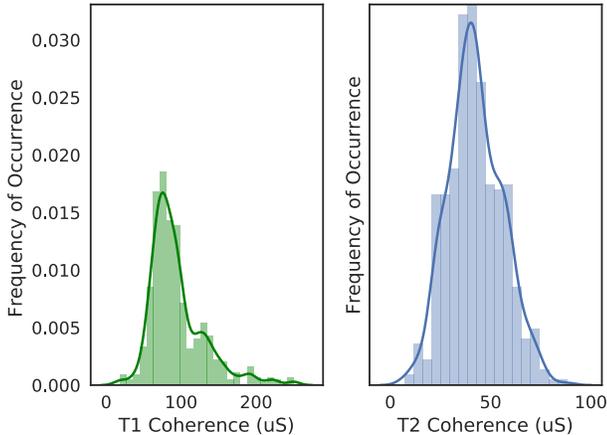

**Fig. 6:** Distribution of (a) T1 Coherence time (b) T2 Coherence time

### B. Distribution of Error-Rate of Single-Qubit Operations

Single qubit operations rotate the quantum state from one point to other on a state-sphere. They are performed by applying a microwave signal with a set duration and frequency on the qubit device. Unfortunately, qubit devices are highly non-linear and a small perturbation in biasing or experimental conditions can cause drift in device characteristics. This can cause variation in robustness of the quantum operations Figure 7 shows the distribution of error-rate for single-qubit operations. The data shows a large fraction of the error-rate below 1%. In general, single-qubit operations are more robust compared to two-qubit operations and the overall error rate of the system usually gets determined by the two-qubit operations.

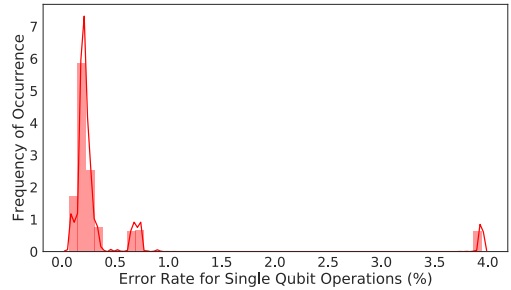

**Fig. 7:** Distribution of the error-rates of single-qubit operations.

### C. Distribution of Error-Rate of Two-Qubit Operations

Two-qubit operations are essential to entangle quantum states and move data around. They are one of the most dominant operations in the NISQ programs. In IBM quantum computers, two-qubit operations are performed by applying microwave pulses on target devices, control qubit devices as well as on the coupling link that connects the two. Similar to single-qubit operations, two-qubit operations suffer from variation in error-rate i.e. there is a fraction of coupling links significantly weaker than most of the links. We monitor the reliability of two-qubit operations for the IBM quantum computer. Figure 8 shows the distribution error-rate of two-qubit operations for the 20 qubit machine. It consists of data from 76 coupling links. The mean error rate for the two-qubit operation is 4.3% and standard-deviation is 3.02%.

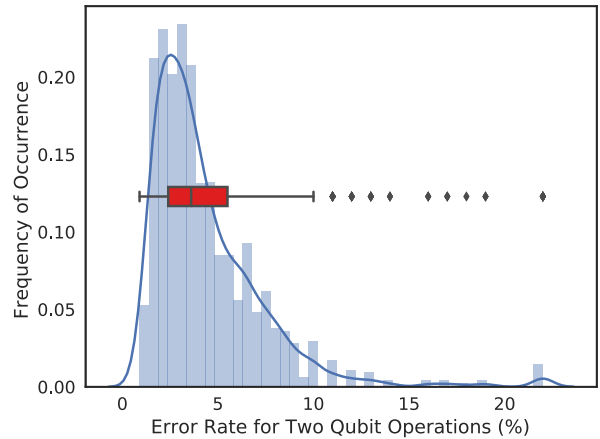

**Fig. 8:** Distribution of the error-rates of two-qubit operations for 76 pair of qubits in IBM's 20 qubit computer.

### D. Temporal Variation in Error-Rate of Two-Qubit Operations

Error-rate of a link can change with time. IBMQ-20 are frequently re-calibrated to ensure that the characterization is reliable. However, a qubit and the associated coupling links can change their behaviour across two different calibration points. For example, a qubit pair with a low error rate on one day can have completely opposite behaviour on the other. This might result from tuning parameters, drifts, local temperature gradient and other experimental factors. Figure 9 shows a time-series of error-rate for three coupling-links. From this data, we observe that error-rate of the links tend to retain their mean error characteristics and stronger links tend to remain strong.

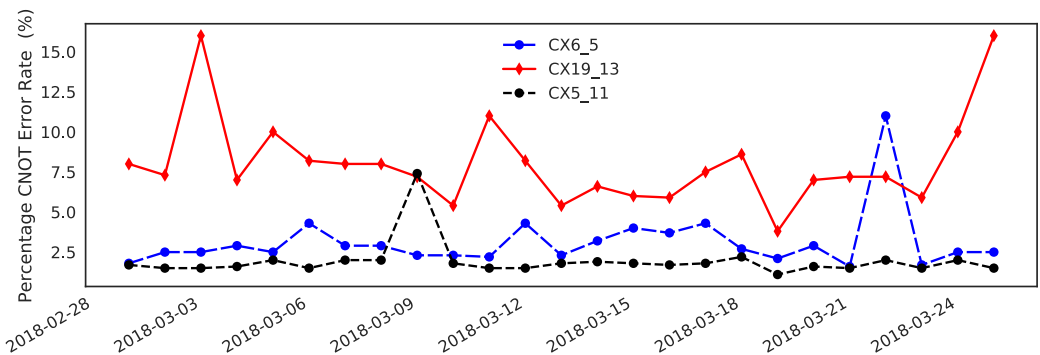

**Fig. 9:** Temporal variation in error rate of two-qubit operations (shown for three links). Note that for most of the calibration periods, the strong links tends to remain strong and the weak link tends to remain weak.

### E. Spatial Variation in Error-Rate of Two-Qubit Operations

Figure 10 shows the layout of the IBM-Q20 qubit computer. Circular nodes represent the qubits and the directed-edges in the graph represent a coupling link that is used for performing two-qubit operation between a pair of qubits. The weight on the edge shows the strength of the link that represents the average probability of failure of the link. For example, a link between $Q0$ and $Q1$ has a probability of failure of 0.04. Note that the link between $Q14$ and $Q18$ has the highest probability of failure (0.15) and there are several links with probability of failure as low as 0.02. Thus, there is a variation of 7.5x between the failure rate of the strongest links versus the weakest link.

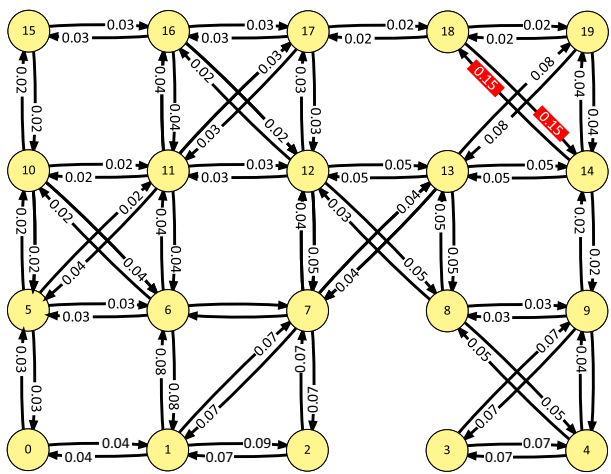

**Fig. 10:** Layout of IBM's 20 qubit machine, each edge represents a possible 2-qubit operation. The label on the edge represent the probability of failure on that link when an operation is performed. The best link(s) have an error-rate of 0.02 and the worst link has 0.15, so a difference in strength of 7.5x.

We observe that for all the metrics we have analyzed (coherence times, error-rate of single qubit operations, and error-rate of two-qubit operations), there is significant variation in the behavior of qubits and links. Given that the data for this variation can be obtained using characterization (which is done periodically anyway), we can use the variation data and develop variation-aware policies. We first define our evaluation methodology and the figure of merit (for assessing system level reliability) and then present our proposals.

## IV. DESIGN METHODOLOGY

To the best of our knowledge, this is the first paper to analyze system-level reliability of NISQ computers. As this area is still in the stage of infancy, there is no established methodology to perform such evaluations. We describe our evaluation infrastructure, present possible figure-of-merits for evaluating system-level reliability, and then provide details of the benchmarks we use for our analysis.

### A. Evaluation Infrastructure

We use the iterative model for NISQ where the same workload is executed a large number of times, and the output is analyzed. To perform our evaluations, we built a Monte-Carlo simulator, as shown in Figure 11. This simulator accepts (a) NISQ program (b) Layout, configuration, and error rate, and (c) management policies. The simulator tracks if the program completed without an error, or if an error happened, how many instructions were completed before the error. We perform 1 million trials for each workload to get reliable estimates.

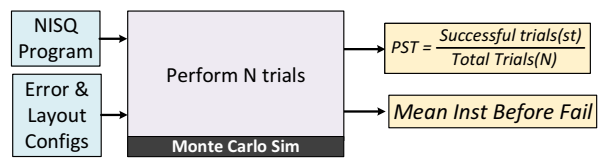

**Fig. 11:** Monte-Carlo simulator for gathering failure statistics of the overall system, for a given program, layout, and error model. We perform 1 million trials for each evaluation.

### B. Figure-of-Merit for System-Level Reliability

The figure of merit for system level reliability depends on the type of workload. For workloads that tend to finish without errors at least some of the times, we can measure what percentage of trials were completed without an error. Our first figure-of-merit is indeed based on this observation. We introduce, *Probability of Successful Trial (PST)*, to denote the probability that a program completes without any errors. $PST$ is the ratio of successful trials to the total trials performed in the Monte-Carlo trials, as shown in Figure 11.

Unfortunately, PST is dependent on both the error distribution, and the length of the program. For example, shorter quantum programs have larger PST, whereas long programs can have PST close to zero. Furthermore, a program that doesn't terminate successfully with high probability would require impractically large number of total trials to get even one successful trial. For such long-running program, that have a negligible likelihood of completing without errors, we define the second figure-of-merit, *Mean Instructions Before Failure (MIBF)*. MIBF is the average number of instructions a program can execute before it encounters a first error. MIBF represents the number of error free operations a NISQ can perform. To calculate MIBF: we run a Monte Carlo trial until an error occurs. When an error occurs we store the number of instructions executed before failure (IBF) and restart the trial for the program. We repeat this process, and compute the average.

*C. Benchmarks*

For our evaluations, we use micro-benchmarks and random-benchmarks used by the prior studies on qubit allocation [9]. These micro-benchmarks are scaled down version of larger quantum applications and subroutines. We classify the benchmarks into two groups: terminating and non-terminating. Benchmarks with less than 0.1% PST are classified as non-terminating programs and the rest are terminating program. For terminating workloads we use PST as the figure-of-merit, and for non-terminating, we use MIBF. Table I shows the 9 benchmarks used in our study, their description, number of quantum instruction performed, and the number of qubits.

**TABLE I:** Benchmark Characteristics

| Benchmark | Description | Q-Instructions | Qubits | Metric |
|---|---|---|---|---|
| alu | Quantum adder | 173 | 20 | PST |
| ising | Ising Model | 790 | 16,20 | PST |
| qft | Quantum Fourier Trans. | 512 | 16,20 | PST |
| cnt35 | Random benchmark | 384 | 16 | PST |
| rd84 | Random benchmark | 1000 | 20 | PST |
| gse | Quantum Chemistry | 39k | 14 | MIBF |
| inc | Q-Arithmatic | 10k | 16 | MIBF |
| dist | Q-Arithmatic | 38k | 16 | MIBF |
| sqrt | Q-Arithmatic | 7k | 13 | MIBF |
| rnd2 | Random benchmark | 28k | 20 | MIBF |

*D. Layout Configuration and Error-Rate Parameters*

The layout configuration specifies the number of qubits and their connectivity. For our studies, we use the IBM-Q20 layout. The error-rate parameters describe the error rates for single-qubit, two-qubit and measurement operations. We model the errors in quantum operations as independent trials with a fixed error rate. The data collected from the IBM-Q20 is used to model the error rate distributions. Unfortunately, existing error rates are high (worst case two qubit error rate is 15%). This restrict the number of instructions that we can run (MIBF ranges form 10 to 80). To understand the effectiveness of different policies, we scale all the error rates down by a factor of 10. In this paper, we limit our focus on the impact of only operational errors and do not analyze the impact of coherence errors (as these errors are not dependent on the instructions performed).

## V. VARIATION-AWARE QUBIT MOVEMENT

The characterization data of IBM-Q20 that was presented in Section III showed significant variability in the error rates of qubits and links. We can use the characterization data to develop *Variation-Aware* policies that can improve the overall system reliability of the NISQ computer. In this section, we look at providing variation-awareness to Qubit-Movement policy.

*A. The Problem of Qubit-Movement*

The Qubit-Movement policy is responsible for deciding the route to take while going from one location to another.[1] Such a policy can consider all possible routes and pick the one that requires the fewest number of SWAP instructions. Fortunately, most of the designs for quantum computers use a mesh-like network, so all the choices that go either in the X direction or Y direction towards the destination will have identical Manhattan distance, and hence identical number of SWAP instructions. For example, for the 6-qubit quantum computer shown in Figure 12, if we want to go from physical qubit A to physical qubit F, all three routes (A-B-C-F, A-D-E-F, A-D-C-F) have identical hop counts (3), and the Qubit-Movement policy can choose any of these routes. It may consider making the Qubit-Movement decision simple by first going in the "X" dimension and then going in the "Y" dimension (or vice versa) – while such a policy would ensure the shortest route (minimum number of SWAP instructions), using such a design would exclude the possibility of selecting the route A-D-C-F.

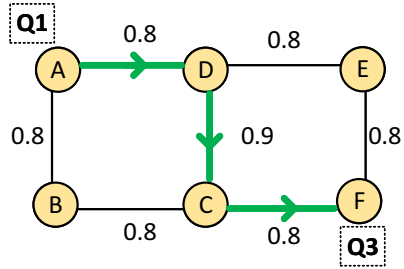

**Fig. 12:** A 6-qubit quantum computer, where each link has a probability of success. To move Q1 (at A) to Q3 (at F), a variation-aware policy would use route A-D-C-F as it maximizes the overall probability of success of the movement

*B. "Variation-Awareness" in Qubit-Movement Policy*

Given that there is variation in the error-rates of different links, policies (such as X-first or Y-first) that simply choose one choice among the list of shortest routes will not always provide the best overall system reliability. For example, the number on each link in Figure 12 shows the probability of success of the link. Route A-D-F would maximize the probability of success of the overall movement from A to F, and a variation-aware policy would choose such a route.

---

[1]Qubit-Movement policy is analogous to network-routing algorithms, which decide the path followed by a packet from the source to destination within a network. Similar to how network-routing algorithms try to minimize the "hop count", Qubit-Movement policies try to minimize the number of SWAP instructions. Network-routing algorithms make localized decisions at each node, so they must be designed carefully to avoid deadlocks. However, Qubit-Movement is orchestrated globally by the compiler, with the knowledge of the utilization of all links, so it is easy to avoid schedules that cause deadlocks.

## C. Design of Variation-Aware Qubit-Movement Policy

We propose *Variation-Aware Qubit Movement (VQM)* that tries to perform Qubit-Movement while taking into account the variation in the per-link error rates. VQM tries to select the paths with highest reliability for the data movement and actively tries to avoid paths that may have poor reliability.

For implementing VQM, we assume that the characterization data of the error rates for different links is available at compile time, and that this characterization data remains valid during the execution of the workload. VQM compiles the application and tries to select the route that tries to maximize system reliability[2]. For selecting the route, VQM simply forms a cost graph where each link has a probability of success, and the overall probability of success of a route is computed as the product of probability of success of the individual links. VQM selects the route that maximizes the probability of success for the overall route. For small networks, a brute force approach for selecting the best route can be used, while for a larger networks, a greedy algorithm can be applied. In our solution, we use Dijkstra's greedy algorithm for finding the path with the lowest cost.

VMQ can select a longer path over the shortest path, if the longer path has higher reliability. This will result in executing extra number of SWAPs in the workload. We can limit this extra SWAP instructions using a parameter *Maximum Additional Hop (MAH)*. VAQ with such limitations will select the path with the lowest cost, that does not have more additional hops than dictated by this parameter. We use $MAH = 4$ to analyze such hop-limited VQM. Note that our system reliability numbers include the failures due to any extra hops added by VQM.

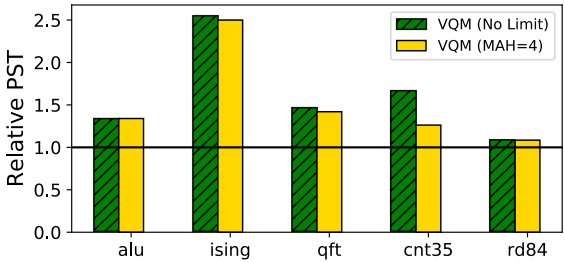

**Fig. 13:** Impact of VQM on the Probability of Successful Trials (PST). Note that reported PST numbers are normalized to PST of the baseline policy that selects the shortest route.

## D. Impact of VQM on Probability of Successful Trials

Figure 13 reports the Relative-PST for five micro-benchmarks when compiled with VQM and the constrained version of VQM. The baseline policy is variation-unaware and is optimized to take the shortest route ignoring the link strengths. VMQ improves the PST for `ising`, `qft`, and `cnt35`, by 2.5x, 1.46x, and 1.66x respectively. In the baseline design, all three programs fail due to a small fraction of unreliable links that has considerably worse error-rate than average. VMQ avoid the weaker links by taking a more robust path. Both `rd84` and `alu`, has considerably large number of SWAP instructions operating over a larger set of qubits.

[2]In conventional computer systems, applications may be compiled once, and run unchanged for several years. However, it is reasonable in NISQ domain to assume that each time the workload is scheduled, it gets recompiled and then repeated trials are performed with the updated executable.

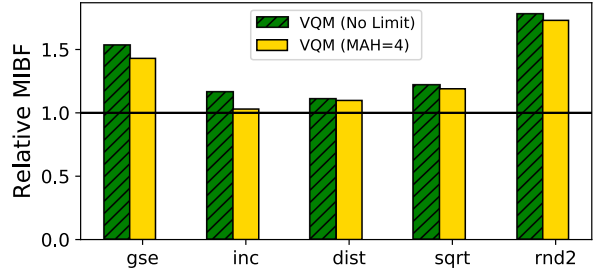

**Fig. 14:** Impact of VQM on the Mean Instructions Before Failure (MIBF). Note that reported MIBF numbers are normalized to MIBF of the baseline policy that selects the shortest route.

## E. Impact of VQM on Mean Instructions Before Failure

For applications that have negligible probability of successful completion, we use *Mean-Instructions Before Failure (MIBF)* as the figure of merit. Figure 14 shows the Relative-MIBF for five benchmarks when compiled with VMQ. VQM improves the MIBF by 10% to 80%. The largest improvement is seen for a random benchmark `rnd2` that attempts to perform small number of long range SWAPs and then performs large number of single qubit operations.

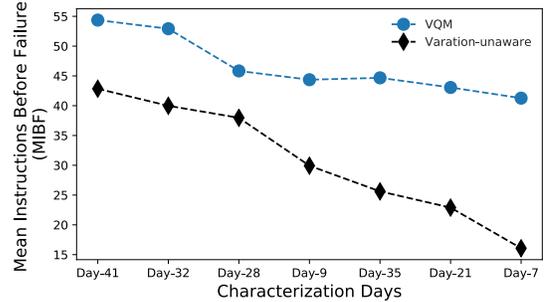

**Fig. 15:** MIBF of the baseline (variation-unaware) and VQM against inter-day variation in error rates. Note that the days on the x-axis are sorted from lowest variation to the highest.

## F. Impact of Inter-Day Variation in Error Rates of IBM-Q20

Our reliability evaluations are done by using the average of the error-rates measured over 52 days, and we scale the average error-rate of each component by 10x (to ensure that at least some of the benchmarks can finish without errors). For this section, we use the raw IBM-Q20 data and show the robustness of VQM to the inter-day variation in error rates. We use `ising` design and recompiled it with the data available for each day. Figure 15 shows the MIBF for IBMQ-20 for VQM and the baseline (variation-unaware) policy. The data is displayed for 7 representative days, sorted from the day with the lowest variation to the day with the highest variation. The MIBF of the variation-unaware policy drops significantly, whereas VQM is robust to the inter-day variation in error rates.

## VI. VARIATION AWARE ALLOCATION

### A. The Problem of Qubit-Allocation

The Qubit-Allocation policy is responsible for assigning the program qubits to the physical qubits [8]. The decisions of the Qubit-Allocation policy affect the data-movement patterns between qubits. For example, if two qubits are placed far away (A and F in Figure 16), they will need large number of SWAP operations. Recent compiler studies have looked at qubit allocation and have proposed algorithms that minimize the number of SWAP instructions [10], [9], [8].

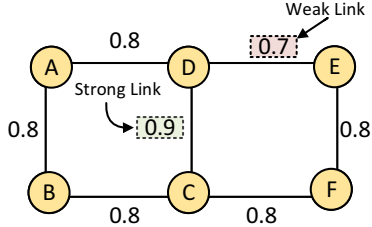

**Fig. 16:** A 6-qubit quantum computer, where each link has a probability of success. If two qubits are to be mapped, a variation unaware policy may map them to D and E (the link with worst error rate), whereas, a variation-aware policy will allocate them to D and C (the link with the best error rate).

### B. "Variation-Awareness" in Qubit-Allocation

Existing policies for qubit allocation are oblivious to the variation in the link reliability. They simply use an allocation that minimizes the number of SWAP instructions. For example, if we want to allocate 2 qubits for the machine in Figure 16, these policies may pick any two neighboring qubits, including D and E, which are connected by the weakest link. If the allocation policy was aware of the variation, it would pick D and C, which are connected by strongest link. To that end, we propose *Varaition-Aware Qubit Allocation (VQA)* policy.

### C. Design of Variation-Aware Qubit-Allocation Policy

The frequently operated qubits have higher probability of failure. To minimize the probability of failure, frequently used qubits are mapped to the strongest qubit block that consists of qubits with highest connectivity and reliable coupling links. We define, connectivity-strength as sum of all the coupling-link success probabilities for a device. For example, D in Figure 16) has 3 links with success probabilities of 0.8, 0.7, and 0.9 resulting in a connectivity strength of 2.4. VQA computes the connectivity strength for all the qubits and select the connected sub-graph which has the maximum total connectivity strength.

VQA also need to balance SWAP count and reliability. Mapping program-qubits to strongest physical qubits without understanding the access pattern of the program lead to extra SWAPs. The extra SWAPs can reduce the system reliability significantly. VQA uses locality aware allocation. When mapping the qubits on the strongest qubit block, VQA computes the SWAP count of a pair of qubits for the first-N instruction in the program, and order the program-qubit to physical qubit mapping based on the SWAP count.

### D. Impact of VQA on Probability of Successful Trials

The potential for improvement with intelligent mapping is greater when there are only a few program qubits to be mapped to a larger number of physical qubits. Several of our workloads try to map 16 to 20 qubits on a 20-qubit machine. To show the potential benefit of VQA, also created smaller version of `ising-10` (10-qubit) and `qft-10` (10-qubit) as these are tunable workloads. Figure 17 shows the relative-PST for the micro-benchmarks normalized to the baseline. VQA is built on top of VQM, so we show results for VQM and VQM+VQA. We observe that VQA can provide improvement for some workloads (`qft`) and this improvement will be greater if the workload had fewer qubits. For example, `qft-10` with 10 qubits shows more improvement than the `qft-16`, and a similar pattern is present for `ising`.

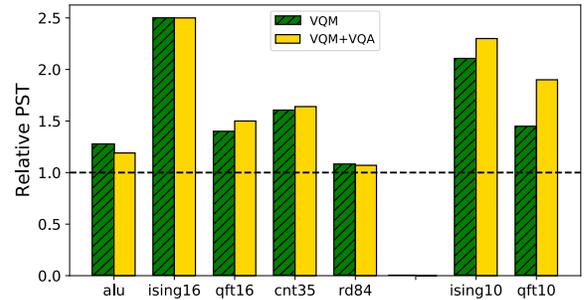

**Fig. 17:** PST for VQM only and VMQ implemented with VQA. The bars on the right are 10-qubit implementations. VQA provides more benefit for workload with fewer qubits.

### E. Impact of VQA on Mean Instructions Before Failure

Figure 18 shows the Normalized MIBF for a system that uses only VQM and for a system that implements both VQM and VQA. The MIBF is normalized to the baseline system. For some workloads (`inc`, `dist`, `sqrt`) VQA provides significant benefits above VQM. However, workloads such as gse and rnd2 get a degradation. Benchmark `gse` has random chain of CNOTs in the initial part of the program that changes the mapping significantly, sometimes leading frequently used qubits to weak links devices. Furthermore, VQA is mapped based on the first few instructions, there are cases where a qubit is dormant for some time and then it becomes active.

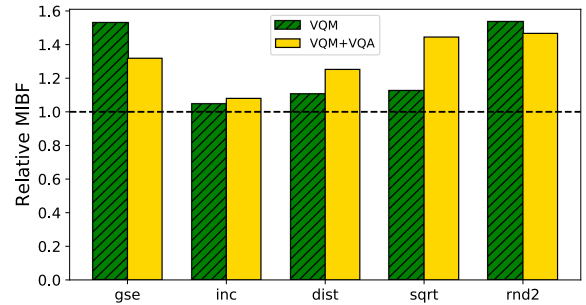

**Fig. 18:** MIBF for VQM only and VMQ implemented with VQA.

## VII. CASE STUDY: PARTITIONING QUANTUM COMPUTER

We have explored the concept of variation-aware policies for Qubit-Movement and Qubit-Allocation. However, this concept can be used to provide insights into other design trade-offs that may come in NISQ systems. We do a case study for a scenario, where the workload requires half or fewer qubits than what is physically available, and the computer can be partitioned to run multiple copies of the same workload (to provide more trials per unit time). We analyze whether it makes sense to partition the NISQ computer or not, in such scenarios.

### A. Choice: Two Weak-Copies versus One Strong-Copy

When the number of program qubits are less than or equal to half of the physical qubits, we can run two copies of the same program. In an ideal world, the simultaneously running two copies can provide twice as many number of error-free trials per unit time. However, for a quantum computer with variation, running two copies restrict the program qubit to physical qubit mappings. For example, running a single copy provide an opportunity to choose the strongest set of qubits and links in a given quantum computer, whereas, running two copies would constrain us to also use weaker set of qubits and links. Thus, the single copy would try to maximize the PST (Probability of Successful Trial) for a given trial, even if it means sacrificing the increased trials per unit time that would be possible with two-copies. Whereas, having two-copies provides more trials per unit time at the expense of PST for each trial. On a given NISQ with variable reliability, should we run two weak copies or run one strong copy of the program?

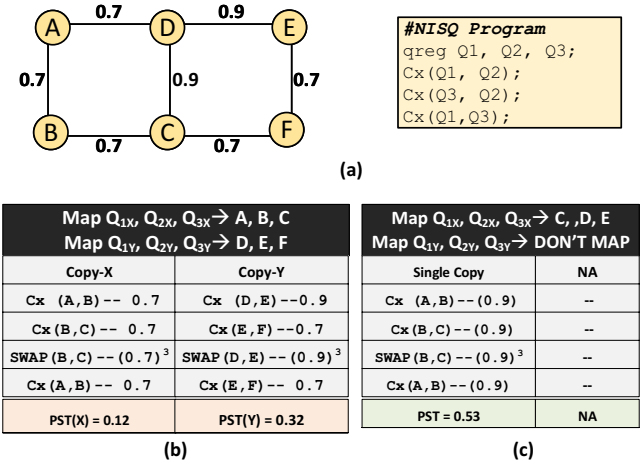

**Fig. 19:** (a) NISQ with six qubits using mesh connectivity. A CNOT reliability is reported on the top of each link. (b) (c) Mapping policy that runs two copies of a NISQ program (d) Mapping policy that runs one copy using the strongest links

Consider a hypothetical quantum computer with six physical qubits with a mesh-layout as shown in the Figure 19(a). The edge-weights in the graph show the strength of the coupling links. For a quantum program with three program qubits as shown in the Figure 19(a), we can either run two copies by partitioning the quantum computer or run just one copy. Figure 19(b), show two copies of a program: Copy-X and Copy-Y running on a quantum computer. The success probability of individual copy can be calculated by multiplying all the success probabilities of operations in the program. For example, Figure 19(b) shows the PST for Copy-X and Copy-Y to be 0.32 and 0.12 respectively. Thus, running two copies does not increase the rate at which successful trials can be done by 2x, instead in our case it is only 37.5% (0.44/0.32).

For the example program, if we choose to run a single copy, we can intelligently select the strongest subset of qubits and links to improve the overall reliability. Figure 19(c) shows one such example whereby choosing to run just one strong copy can improve the cumulative PST. When running two copies, the constraints on connectivity restricts the use of link CD which is one of the strongest links. When running two copies, programmer has to resort to the weaker links. Whereas, when running a single copy, we can pick most reliable links and achieve better PST as shown in the Figure 19(b).

### B. Benchmark-Based Evaluation

We extend our simulation infrastructure to support two copies of the same workload. For the two-copy mode, we explore all possible partitions and select the best. Note that besides the number of copies, movement and the mapping algorithm used for both of the policies are identical. The only difference is the available number of qubits. For the evaluation in this section, we use the figure of merit as *Number of Success Trials Per Unit Time (STPT)*, as it captures both the PST and the increased rate of trials with two copies. For the benchmarks, we only use the ones for which we can analyze the PST (as benchmarks that use MIBF do not finish, running multiple copies will not be helpful). We modify these benchmarks to use only 10 qubits. Figure 20 shows the STPT of the single strong-copy and two-copies, both normalized to the STPT of the two-copies. For this study we selected the three terminating workloads that can operate with ten qubits. We observe that sometimes two-copy is better (ising) and sometimes one strong-copy is better (qft). So, a user can leverage our analysis to estimate which solution is likely to perform better for the workload and used that solution. Thus, our variation-aware policies may be useful in enabling *Adaptive Partitioning* for NISQ, where the decision between one strong copy versus two-copies can be based on STPT.

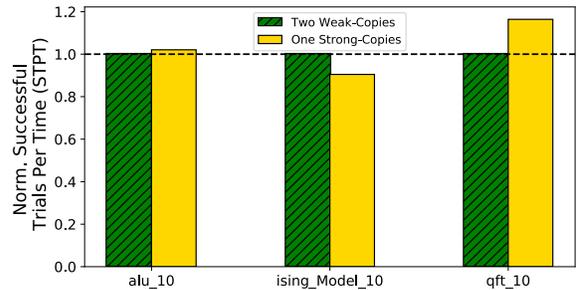

**Fig. 20:** Successful Trials per Unit Time (STPT) when running (a) two-copies (b) one strong copy. Note that the micro-benchmarks were modified to have 10 program qubits.

## VIII. DISCUSSION OF LIMITATIONS

Computer architecture for NISQ-Era quantum computers is still in the stage of infancy. There is no established evaluation infrastructure to estimate the impact of different policies on the system-level reliability of quantum computers using benchmarks or applications. Our paper undertakes this task, however, as with any initial research, it is based on a large number of assumptions, which may not hold, as the technology matures. We discuss some of the limitations of our study:

**Evaluation Infrastructure:** Our evaluations are based on the iterative model for NISQ computers. However, not all NISQ applications will use such a model. For these applications, *Probability of Successful Trial (PST)* is not a meaningful metric.

**Workloads:** Our evaluations are done using small kernels and random benchmarks, similar to the ones used in the area of compilation for quantum computers. These kernels and benchmarks may not be representative of the emerging NISQ applications that may get developed over the coming years.

**Error Models:** We make several simplifying assumptions such as no-correlations between errors, static error-rates, and ignore retention errors. Noise in the real-world quantum computer can be significantly complex and difficult to model accurately.

The basic insight in our work is that there is variation in reliability of different components. Exploiting the variability allows better-than-worst-case behavior and avoids the overall system reliability getting dictated by a few weak components. While we expect this basic insight to be useful for future quantum computers, some of our assumptions about evaluations and error models may get redefined as the field progresses.

## IX. RELATED WORK

Our work spans a large number of technical areas: Quantum computing, system-level reliability, compilation and register allocation policies, and network-routing policies.

Early works in quantum system architecture provided a blueprint for quantum systems by defining system abstractions [15], [16], [17], [18], [19], [20]. Prior work on quantum instruction set architecture, microarchitectural primitives has built strong foundations for quantum system architects. A large body of work has also focused on compiler level problems for quantum computers and quantum compilers have provided means to synthesize, simulate, and analyze quantum programs [21], [22].

Recent, experimental breakthroughs and large investments in building small scale quantum computers has encouraged system architects to focus on small-scale quantum system designs. To this end, several ideas on architecting superconducting systems has been proposed [23], [24]. Furthermore, the availability of 20 qubit machine to general public has sparked the interest in building compilers and assembler tools for the near-term quantum computers. And recent works provide good theoretical understanding on the mapping problems [10]. Furthermore, researchers have built the heuristics that can be used to solve this problem in optimal time [25]. Along with general mapping problems, researcher have started focusing on the machine specific mapping problems. To this end, the compiler tools are built for the IBM 5 and 16 qubit machine respectively [9], [8].

Recently IBM researchers proposed the *Quantum Volume (QV)* metric to compare quantum computers with different number of qubits and varying degree of connectivity. However, this metric does not capture the reliability loss due to variation, is an application-agnostic metric, and does not account for policy decisions in quantum computers. Whereas, our proposed metrics of PST and MIBF denote the system failure rate for a given system (with a fixed number of qubits and connectivity) is designed to quantify the reliability impact of device variation, policy decisions, and benchmark-dependent behavior.

The problem of data movement in quantum computers is similar to routing in on-chip networks and our proposal is similar in spirit to the problem of routing data with faulty network links [26], [27], [28], [29], [30].

## X. SUMMARY

The availability of small-scale quantum computers has provided an unique opportunity to computer system researchers to understand and solve the problems that occur in operating a realistic quantum computer. In this paper, we study the policies for *Qubit-Allocation* (mapping of program qubits to machine qubits) and *Qubit-Movement* (routing qubits from one location to another to perform entanglement) for available quantum computers. Given that existing quantum computers use restricted connectivity, these policies have a significant impact on the operations required for two qubits to communicate. We observe that there can be variation in the error rates of different qubits and links, which can mean that prior studies that try to minimize communication may not maximize overall system reliability. The system reliability of quantum computers can be increased by steering more operations towards stronger qubits and links, and less operations towards weaker qubits and links. To this end, our paper makes the following contributions:

- We present the characterization data for the IBM-Q20 quantum computer for 52 days. This data shows that there is significant variation in the error rate of qubits and links.
- We develop an evaluation methodology to assess the impact of device variation and management policies on the system-level reliability of quantum computers. We define two metrics (PST and MIBF) for quantifying reliability.
- We propose *Variation-Aware Qubit Movement* policy that takes variation of the links into account, and tries to pick a route that has the lowest probability of failure.
- We propose *Variation-Aware Qubit Allocation* policy that maps the data such that the program qubits are located to use the stongest links, and avoid the weaker links.

Our models can also helps in understanding the resource sharing and partition problems in the existing and near-term quantum computers, such as deciding between running one strong-copy versus two concurrently running copies, for applications that require few qubits. As the domain of quantum computing moves from theory, to devices, to realistic systems, it is important to have studies that make it easier for computer architecture to reason about and optimize existing and future quantum computers. Our paper takes a step in this direction.

## REFERENCES

[1] P. W. Shor, "Polynomial-time algorithms for prime factorization and discrete logarithms on a quantum computer," *SIAM review*, vol. 41, 1999.

[2] M. B. Hastings, D. Wecker, B. Bauer, and M. Troyer, "Improving quantum algorithms for quantum chemistry," *Quantum Info. Comput.*, vol. 15, Jan. 2015. [Online]. Available: http://dl.acm.org/citation.cfm?id=2685188.2685189

[3] W. Knight, "IBM Raises the Bar with a 50-Qubit Quantum Computer," https://www.technologyreview.com/s/609451/ibm-raises-the-bar-with-a-50-qubit-quantum-computer/, 2017, [Online; accessed 3-April-2018].

[4] J. Kelly, "A Preview of Bristlecone, Googles New Quantum Processor," https://research.googleblog.com/2018/03/a-preview-of-bristlecone-googles-new.html, 2018, [Online; accessed 3-April-2018].

[5] J. Hsu, "CES:Intel's 49-Qubit Chip Shoots for Quantum Supremacy," https://spectrum.ieee.org/tech-talk/computing/hardware/intels-49qubit-chip-aims-for-quantum-supremacy, 2017, [Online; accessed 3-April-2018].

[6] J. Preskill, "Quantum computing in the nisq era and beyond," *arXiv preprint arXiv:1801.00862*, 2018.

[7] T. Häner, D. S. Steiger, K. Svore, and M. Troyer, "A software methodology for compiling quantum programs," *arXiv preprint arXiv:1604.01401*, 2016.

[8] M. Siraichi, V. F. Dos Santos, S. Collange, and F. M. Q. Pereira, "Qubit allocation," in *CGO 2018-IEEE/ACM International Symposium on Code Generation and Optimization*, 2018.

[9] A. Zulehner, A. Paler, and R. Wille, "Efficient mapping of quantum circuits to the ibm qx architectures," *arXiv preprint arXiv:1712.04722*, 2017.

[10] A. Shafaei, M. Saeedi, and M. Pedram, "Optimization of quantum circuits for interaction distance in linear nearest neighbor architectures," in *Proceedings of the 50th Annual Design Automation Conference*. ACM, 2013.

[11] T. S. Metodi, A. I. Faruque, and F. T. Chong, "Quantum computing for computer architects," *Synthesis Lectures on Computer Architecture*, vol. 6, 2011.

[12] M. H. Devoret and R. J. Schoelkopf, "Superconducting circuits for quantum information: an outlook," *Science*, vol. 339, 2013.

[13] E. Knill, D. Leibfried, R. Reichle, J. Britton, R. Blakestad, J. Jost, C. Langer, R. Ozeri, S. Seidelin, and D. J. Wineland, "Randomized benchmarking of quantum gates," *Physical Review A*, vol. 77, 2008.

[14] I. B. M. Corporation, "Universal Quantum Computer Development at IBM:," http://research.ibm.com/ibm-q/research/, 2017, [Online; accessed 3-April-2017].

[15] S. Balensiefer, L. Kregor-Stickles, and M. Oskin, "An evaluation framework and instruction set architecture for ion-trap based quantum micro-architectures," in *ACM SIGARCH Computer Architecture News*, vol. 33, no. 2. IEEE Computer Society, 2005.

[16] R. Van Meter and C. Horsman, "A blueprint for building a quantum computer," *Commun. ACM*, vol. 56, Oct. 2013. [Online]. Available: http://doi.acm.org/10.1145/2494568

[17] N. C. Jones, R. Van Meter, A. G. Fowler, P. L. McMahon, J. Kim, T. D. Ladd, and Y. Yamamoto, "Layered architecture for quantum computing," *Physical Review X*, vol. 2, 2012.

[18] K. M. Svore, A. V. Aho, A. W. Cross, I. Chuang, and I. L. Markov, "A layered software architecture for quantum computing design tools," *Computer*, vol. 39, 2006.

[19] N. Isailovic, M. Whitney, Y. Patel, and J. Kubiatowicz, "Running a quantum circuit at the speed of data," in *ACM SIGARCH Computer Architecture News*, vol. 36, no. 3. IEEE Computer Society, 2008.

[20] M. Oskin, F. T. Chong, and I. L. Chuang, "A practical architecture for reliable quantum computers," *Computer*, vol. 35, 2002.

[21] A. JavadiAbhari, S. Patil, D. Kudrow, J. Heckey, A. Lvov, F. T. Chong, and M. Martonosi, "Scaffcc: Scalable compilation and analysis of quantum programs," *Parallel Computing*, vol. 45, 2015.

[22] D. Kudrow, K. Bier, Z. Deng, D. Franklin, Y. Tomita, K. R. Brown, and F. T. Chong, "Quantum rotations: a case study in static and dynamic machine-code generation for quantum computers," in *ACM SIGARCH Computer Architecture News*, vol. 41, no. 3. ACM, 2013.

[23] X. Fu, M. A. Rol, C. C. Bultink, J. van Someren, N. Khammassi, I. Ashraf, R. F. L. Vermeulen, J. C. de Sterke, W. J. Vlothuizen, R. N. Schouten, C. G. Almudever, L. DiCarlo, and K. Bertels, "An experimental microarchitecture for a superconducting quantum processor," in *Proceedings of the 50th Annual IEEE/ACM International Symposium on Microarchitecture*, ser. MICRO-50 '17. New York, NY, USA: ACM, 2017. [Online]. Available: http://doi.acm.org/10.1145/3123939.3123952

[24] X. Fu, L. Riesebos, L. Lao, C. G. Almudever, F. Sebastiano, R. Versluis, E. Charbon, and K. Bertels, "A heterogeneous quantum computer architecture," in *Proceedings of the ACM International Conference on Computing Frontiers*. ACM, 2016.

[25] D. Venturelli, M. Do, E. Rieffel, and J. Frank, "Compiling quantum circuits to realistic hardware architectures using temporal planners," *Quantum Science and Technology*, vol. 3, 2018.

[26] A. DeOrio, D. Fick, V. Bertacco, D. Sylvester, D. Blaauw, J. Hu, and G. Chen, "A reliable routing architecture and algorithm for nocs," *IEEE Transactions on Computer-Aided Design of Integrated Circuits and Systems*, vol. 31, 2012.

[27] D. Fick, A. DeOrio, G. Chen, V. Bertacco, D. Sylvester, and D. Blaauw, "A highly resilient routing algorithm for fault-tolerant nocs," in *Proceedings of the Conference on Design, Automation and Test in Europe*. European Design and Automation Association, 2009.

[28] J. Flich, A. Mejia, P. Lopez, and J. Duato, "Region-based routing: An efficient routing mechanism to tackle unreliable hardware in network on chips," in *Networks-on-Chip, 2007. NOCS 2007. First International Symposium on*. IEEE, 2007.

[29] D. Fick, A. DeOrio, J. Hu, V. Bertacco, D. Blaauw, and D. Sylvester, "Vicis: a reliable network for unreliable silicon," in *Proceedings of the 46th Annual Design Automation Conference*. ACM, 2009.

[30] B. Gavish and I. Neuman, "Routing in a network with unreliable components," *IEEE Transactions on Communications*, vol. 40, 1992.